\documentclass{mn2e}
\usepackage{graphics}

\begin{document}

\title[Star Formation at $z\approx 6$ in the UDF]
{The Star Formation Rate of the Universe at $z\approx 6$ from the
Hubble Ultra Deep Field}
\author[Bunker et al.\ ]{Andrew J.\ Bunker\,$^{1,2}$,
Elizabeth R.\ Stanway\,$^{2}$,
Richard S.\ Ellis\,$^{3}$,
\newauthor
Richard G.\ McMahon\,$^{2}$ \\
$^{1}$\,School of Physics, University of Exeter, Stocker Road, Exeter, EX4 4QL, U.K.\\ {\tt email:
bunker@astro.ex.ac.uk}\\
$^{2}$\,Institute of Astrophysics, University of Cambridge,
Madingley Road, Cambridge, CB3\,0HA, U.K.\\
$^{3}$\,California Institute of
Technology, Mail Stop 169-327, Pasadena, CA~91109, U.S.A.
}
\date{Accepted for publication in MNRAS}

\maketitle

\begin{abstract}
  We determine the abundance of $i'$-band drop-outs in the
  recently-released {\em HST}/ACS Hubble Ultra Deep Field (UDF). Since
  the majority of these sources are likely to be $z\approx 6$ galaxies
  whose flux decrement between the F775W $i'$-band and F850LP
  $z'$-band arises from Lyman-alpha absorption, the number of detected
  candidates provides a valuable upper limit to the unextincted star
  formation rate at this redshift. We demonstrate that the increased
  depth of UDF enables us to reach an $8\,\sigma$ limiting magnitude
  of $z'_{AB}=28.5$ (equivalent to $1.5\,h^{-2}_{70}\,M_{\odot}\,{\rm
    yr}^{-1}$ at $z=6.1$, or $0.1\,L^{*}_{UV}$ for the $z\approx 3$
  $U$-drop population), permitting us to address earlier ambiguities
  arising from the unobserved form of the luminosity function. We
  identify 54 galaxies (and only one star) at $z'_{AB}<28.5$ with
  $(i'-z')_{AB}>1.3$ over the deepest 11\,arcmin$^{2}$ portion of the
  UDF field. The characteristic luminosity ($L^*$) is consistent with
  values observed at $z\approx 3$. The faint end slope ($\alpha$) is
  less well constrained, but is consistent with only modest evolution.
  The main change appears to be in the number density ($\Phi^{*}$).
  Specifically, and regardless of possible contamination from cool
  stars and lower redshift sources, the UDF data support our previous
  result that the star formation rate at $z\approx 6$ was
  approximately $\times 6$ {\em less} than at $z\approx 3$ (Stanway,
  Bunker \& McMahon 2003). This declining comoving star formation rate
  ($0.005\,h_{70}\,M_{\odot}\,{\rm yr}^{-1}\,{\rm Mpc}^{-3}$ at
  $z\approx 6$ at $L_{UV}>0.1\,L^{*}$ for a salpeter IMF) poses an
  interesting challenge for models which suggest that $L_{UV}>0.1\,L^{*}$
  star forming galaxies at $z\simeq 6$ reionized the universe. The
  short-fall in ionizing photons might be alleviated by galaxies
  fainter than our limit, or a radically different IMF.
  Alternatively, the bulk of reionization might have occurred at
  $z\gg6$.
\end{abstract}
\begin{keywords}
galaxies: evolution --
galaxies: formation --
galaxies: starburst --
galaxies: individual: SBM03\#1 --
galaxies: high redshift --
ultraviolet: galaxies
\end{keywords}

\section{Introduction}
\label{sec:intro}

There has been considerable progress over the past decade in
locating galaxies and QSOs at high redshifts. These sources have
enabled us to probe the Universe at early epochs where its
physical characteristics are fundamentally different from those at
the present epoch. Observations of the most distant $z>6.2$ QSOs
(Becker et al.\ 2001, Fan et al.\ 2002) show near-complete
absorption at wavelengths shortward of Lyman-$\alpha$ (Gunn \&
Peterson 1965), suggesting an optical depth in this line that
implies a smooth neutral hydrogen fraction which is increasing rapidly
with redshift at this epoch. Temperature-polarization
cross-correlations in the cosmic microwave background from  {\em
WMAP} indicate that the Universe was completely neutral at
redshifts of $z> 10$ (Kogut et al.\ 2003).

Although there is a growing consensus that cosmic reionization
occurred in the redshift interval $6<z<15$, a second key question is
the nature of the sources responsible for this landmark event.
Optical and X-ray studies to $z\simeq 6$ suggest the abundance of
active galactic nuclei (AGN) at early epochs is insufficient when
account is taken of the relevant unresolved backgrounds (Barger et
al.\ 2003). A more promising source is star-forming galaxies whose
early ancestors may be small and numerous. Along with the escape
fraction for the ionizing photons from the massive and short-lived OB
stars in such sources, a major observational quest in this respect is
a determination of the global star formation rate at early epochs.

In previous papers, our group has extended the {\it Lyman-break}
technique (Steidel, Pettini \& Hamilton 1995; Steidel et al.\ 1996) to
address this question. Using the Advanced Camera for Surveys (ACS,
Ford et al.\ 2002) on the Hubble Space Telescope ({\em HST}) with the
sharp-sided SDSS F775W ($i'$) and F850LP ($z'$) filters, we located
``$i$-drop'' candidates with $z_{AB}'<$25.6 at $z\simeq 6$ for further
study. In a series of papers, we have shown that this selection
technique can effectively locate $z>5.7$ galaxies using ACS images
from the {\em HST} Treasury ``Great Observatory Origins Deep Survey''
(GOODS; Giavalisco \& Dickinson 2002). On the basis of GOODS-South
photometric catalogues published by Stanway, Bunker \& McMahon (2003,
hereafter Paper I), spectroscopic follow-up using Keck/DEIMOS and
Gemini/GMOS field demonstrated our ability to find high redshift
galaxies (Bunker et al.\ 2003, hereafter Paper II; Stanway et al.\ 
2004a). To address potential cosmic variance issues, we performed a
similar analysis in the GOODS-North field, which yielded a consistent
estimate of the surface density of $z\simeq 6$ star forming sources
(Stanway et al.\ 2004b, hereafter Paper III).

Although our initial study (Papers I-III) revealed the importance of
ascertaining the difficult spectroscopic verifications, and
highlighted the problem of contamination from Galactic stars, we
nonetheless determined that the abundance of confirmed star forming
galaxies at $z\sim 6$ must be less than that expected on the basis of
no evolution from the well-studied $z\sim 3-4$ Lyman break population
(Steidel et al.\ 1999).  Working at the robustly-detected bright end
of the luminosity function, in Paper~I we showed that the comoving
star formation density in galaxies with $z'_{AB}<25.6$ is $\approx
6\times$ {\em less} at $z\approx 6$ than at $z\approx 3$.  Our
$z'_{AB}<25.6$ flux limit corresponds to
$>15\,h^{-2}_{70}\,M_{\odot}\,{\rm yr}^{-1}$ at $z=5.9$, equivalent to
$L^{*}_{UV}$ at $z\approx 3$. In Papers I--III we restricted our
analysis to luminous galaxies (where we take ``luminous'' to mean
$L>L*$ for the rest-UV). Other groups have claimed less dramatic
evolution or even no evolution in the volume-averaged star formation
rate, based on the same fields (Giavalisco et al.\ 2004; Dickinson et
al.\ 2004) and similar {\em HST}/ACS data sets (Bouwens et al.\ 2003;
Yan, Windhorst \& Cohen 2003), but these groups work closer to the detection
limit of the images and introduce large completeness corrections for
the faint source counts. The major uncertainty in converting the
abundance of our spectroscopically-confirmed sample in the GOODS
fields into a $z\simeq 6$ comoving star formation rate is the form of
the luminosity function for faint, unobserved, sources. As discussed
in Paper~III, if the faint end of the luminosity function at $z\simeq
6$ was steeper than that at lower redshift, or if $L^{*}$ was
significantly fainter, a non-evolving star formation history could
perhaps still be retrieved.

The public availability of the Hubble Ultra Deep Field (UDF; Beckwith,
Somerville \& Stiavelli 2003) enables us to address this outstanding
uncertainty. By pushing the counts and the inferred luminosity
function of $i'$-band drop-outs at $z\approx 6$ to a limiting lower luminosity
equivalent to one well below $L^{*}_{3}$ for the $z\approx 3$ population,
it is possible to refine the integrated star formation rate at
$z\approx 6$. In this paper we set out to undertake the first
photometric analysis of $i'$-drops in the UDF. Our primary goal is to
understand the abundance of fainter objects with characteristics
equivalent to those of $z\simeq 6$ sources and address uncertainties in
the global star formation rate at this redshift.

The structure of the paper is as follows. In Section~\ref{sec:obs} we describe
the imaging data, the construction of our catalogues and our
$i'$-drop selection. In Section~\ref{sec:discuss} we discuss the
luminosity function of star-forming sources, likely contamination
on the basis of earlier spectroscopic work, and estimate the
density of star formation at $z\approx 6$. Our conclusions are
presented in Section~\ref{sec:concs}. Throughout we adopt the
standard ``concordance'' cosmology of $\Omega_M=0.3$,
$\Omega_{\Lambda}=0.7$, and use $h_{70}=H_0/70\,{\rm
km\,s^{-1}\,Mpc^{-1}}$. All magnitudes are on the $AB$ system (Oke
\& Gunn 1983).

\section{HST Imaging: Observations and $I$-Drop Selection}
\label{sec:obs}

\subsection{{\em HST}/ACS Observations}
\label{sec:HSTobs}

The Hubble Ultra Deep Field (UDF) is a public {\em HST} survey made
possible by Cycle 12 STScI Director's Discretionary Time programme
GO/DD-9978 executed over September 2003 -- January 2004.  For the
present program, the {\em HST} has imaged a single ACS Wide Field
Camera (WFC) tile (11.5\,arcmin$^{2}$) for 400 orbits in 4 broad-band
filters (F435W $B$-band for 56 orbits; F606W $V$-band for 56 orbits;
F775W $i'$-band for 144 orbits; F850LP $z'$-band for 144 orbits). The
UDF field lies within the Chandra Deep Field South (CDF-S) with
coordinates RA=$03^{h}32^{m}39\fs0$, Decl.=$-27^{\circ}47'29\farcs1$
(J2000). As the UDF represents the deepest set of images yet taken,
significantly deeper than the $I$-band exposures of the Hubble Deep
Fields (Williams et al.  1996; 1998), and adds the longer-wavelength
$z'$-band, it is uniquely suited to the goals of our program.

The WFC on ACS has a field of $202''\times 202''$, and a pixel
scale of $0.05''$. The UDF lies within the survey area of
GOODS-South area (Giavalisco et al.\ 2004), surveyed using ACS
with the same filters to shallower depth (3,2.5,2.5 \& 5 orbits in
the $B$, $V$, $i'$ \& $z'$ bands). The UDF was observed at two
main orientations differing by 90 degrees, and within each of
these data was taken in 2 blocks rotated by 4\,deg (orientations
of 310,314,40\,\&\,44\,deg). A 4-point dither box spanning
0.3\,arcsec was used, with half-pixel centres to improve the
sampling. During each ``visit'', there were 3 larger 3\,arcsec
dithers to span the WFC inter-chip gap.

For our analysis we use the reduced UDF data v1.0 made public by STScI
on 09 March 2004.  The pipeline reduction involved bias/dark current
subtraction, flat-fielding, and the combination of background-subtracted
frames rejecting cosmic ray strikes and chip defects. The resulting
reduced images had been ``drizzled'' (Fruchter \& Hook 2002) using the
``MultiDrizzle'' software (Koekemoer et al.\ 2004) on to a finer pixel
scale of $0.03''$, to correct for geometric distortion and to improve
the sampling of the point spread function (PSF). The UDF data has been
placed on the same astrometric system as the GOODSv1.0 images of the
UDF\footnote{Available from {\tt
ftp://archive.stsci.edu/pub/hlsp/goods/v1}}. The photometric zeropoints
adopted were those provided by STScI for the UDF v1.0 data release:
25.673, 26.486, 25.654 \& 24.862 for the $B$, $V$, $i'$ \& $z'$ filters,
where ${\rm mag}_{AB} = {\rm zeropoint}-2.5\log_{10}({\rm counts/s})$.
We have corrected for the small amount of foreground Galactic extinction
toward the CDFS using the {\it COBE}/DIRBE \& {\it IRAS}/ISSA dust maps
of Schlegel, Finkbeiner \& Davis (1998). The optical reddening is
$E(B-V)=0.008$, equivalent to extinctions of $A_{F775}=0.017$ \&
$A_{F850LP}=0.012$.

\subsection{Construction of Catalogues}
\label{sec:cats}

Candidate selection for all objects in the field was performed using
version 2.3.2 of the SExtractor photometry package (Bertin \& Arnouts
1996). As we are searching specifically for objects which are only
securely detected in $z'$, with minimal flux in the $i'$-band, fixed
circular apertures $0\farcs5$ in diameter were trained in the
$z'$-image and the identified apertures used to measure the flux at
the same spatial location in the $i'$-band image by running SExtractor
in two-image mode. For object identification, we adopted a limit of at
least 5 contiguous pixels above a threshold of $2\sigma$ per pixel
($0.0005\,{\rm counts/pixel/s}$) on the data drizzled to a scale of
0\farcs03~pixel$^{-1}$. This cut enabled us to detect all significant
sources and a number of spurious detections close to the noise limit.
As high redshift galaxies in the rest-UV are known to be compact
(e.g., Ferguson et al.\ 2004, Bremer et al.\ 2004, Bouwens et al.\ 
2004), we corrected the aperture magnitudes to approximate total
magnitudes through a fixed aperture correction, determined from bright
compact sources: $-0.11$\,mag in $i'$-band and $-0.14$\,mag in
$z'$-band, the larger latter correction arising from the more extended
PSF wings of the $z'$-band.

The measured noise in the drizzled images underestimates the true
noise as adjacent pixels are correlated.  To assess the true
detection limit and noise properties, we examined the raw ACS/WFC
images from the {\em HST} archive and measured the noise in
statistically-independent pixels.  For the 144-orbit $z'$-band, we
determine that the $8\,\sigma$ detection limit is 
$z'_{AB}=28.5$ for our $0\farcs5$-diameter aperture.  This is
consistent with the noise decreasing as $\sqrt{\rm time}$ from the
5-orbit GOODSv1.0 to the 144-orbit UDF $z'$-band. We adopt this
high $S/N=8$ cut as our conservative sample limit.
We trimmed the outermost edges where fewer frames overlapped in order
to exploit the deepest UDF region, corresponding to a survey area of
11\,arcmin$^{2}$.  From the output of SExtractor 
we created a
sub-catalogue of all real objects brighter than $z'_{AB}<28.5$\,mag
($8\,\sigma$ in a $0\farcs5$-diameter aperture), of which 63 appear to
be promising $i'$-band dropouts (see $\S$2.3) with $(i'-z')_{AB}>1.3$.

\begin{figure}
\resizebox{0.98\columnwidth}{!}{\includegraphics{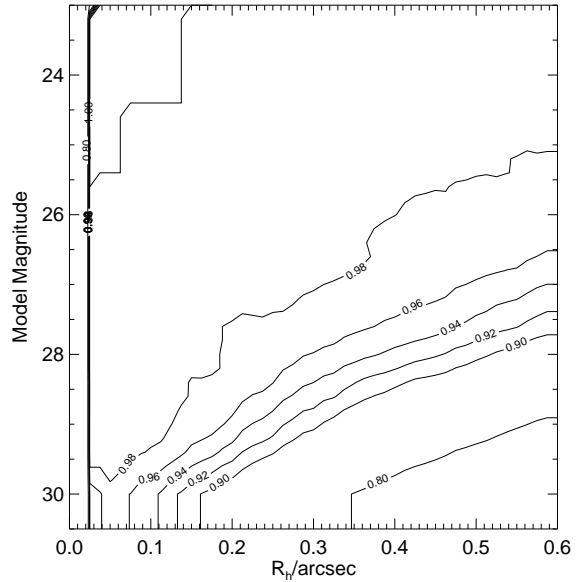}}
\caption{The completeness (normalized to unity) for artificial
galaxies added to the UDF $z'$-band image, as a function
of total magnitude and half-light radius; we re-ran SExtractor
on this image to assess the fraction of artificial galaxies recovered.
The completeness is $>97$\% for $R_h<0\farcs2$ and $z'_{AB}<28.5$.}
\label{fig:completeness}
\end{figure}

To quantify possible incompleteness in this catalogue, we adopted two
approaches. First we examined the recovery rate of artificial galaxies
created with a range of total magnitudes and sizes. We used de
Vaucouleurs $r^{1/4}$ and exponential disk profiles, convolved with
the ACS/WFC PSF derived from unsaturated stars in the UDF images.
Secondly we created fainter realisations of the brightest $i'$-dropout
in the UDF confirmed to be at high redshift (SBM03\#1 with
$z'_{AB}=25.4$, confirmed spectroscopically to be at $z=5.83$ by
Stanway et al.\ 2004b; Dickinson et al.\ 2004). By excising a small
region around this $i'$-dropout, scaling the sub-image to a fainter
magnitude, and adding it back into the UDF data at random locations,
we assessed the recoverability as a function of brightness. For such
objects we recover 98\% of the simulated sources to $z'_{AB}=28.5$,
the remainder being mainly lost via source confusion through
overlapping objects. From these analyses, we determine that, for
unresolved sources ($r_{h}=0\farcs05$), we are complete at our
$8\,\sigma$ limit of $z'_{AB}=28.5$, and are 97\% complete at this
magnitude for $r_{h}=0\farcs2$ (Figure~\ref{fig:completeness}). For
objects with larger half-light radii we will underestimate the
$z'$-band flux due to our $0\farcs5$-diameter photometric aperture.
However, this effect is small for our sample of compact sources
(Table~1 lists both the $0\farcs 5$-diameter magnitudes with an
aperture correction which we adopt, and the SExtractor ``MAG\_AUTO''
estimate of the total magnitude using a curve-of-growth: these are
broadly consistent).

At the relatively bright cut of $z'_{AB}<25.6$ used in Paper~I from the
GOODSv0.5 individual epochs, the UDF data is 98\% complete for sources
as extended as $r_{h}=0.5$\,arcsec. Interestingly, we detect no extended
(low surface brightness) $i'$-drops to this magnitude limit in addition
to SBM03\#1 (Papers~I,III) in the deeper UDF data. This supports our
assertion (Paper~I) that the $i'$-drop population is predominantly
compact and there cannot be a large completeness correction arising from
extended objects (c.f.\ Lanzetta et al. 2002). The ACS imaging is of
course picking out H{\scriptsize II} star forming regions, and these
UV-bright knots of star formation are typically $<1$\,kpc ($<0\farcs 2$
at $z\approx 6$) even within large galaxies at low redshift.

\subsection{$z\approx 6$ Candidate Selection}
\label{sec:RedshfitDiscrim}

\begin{figure}
\resizebox{0.48\textwidth}{!}{\includegraphics{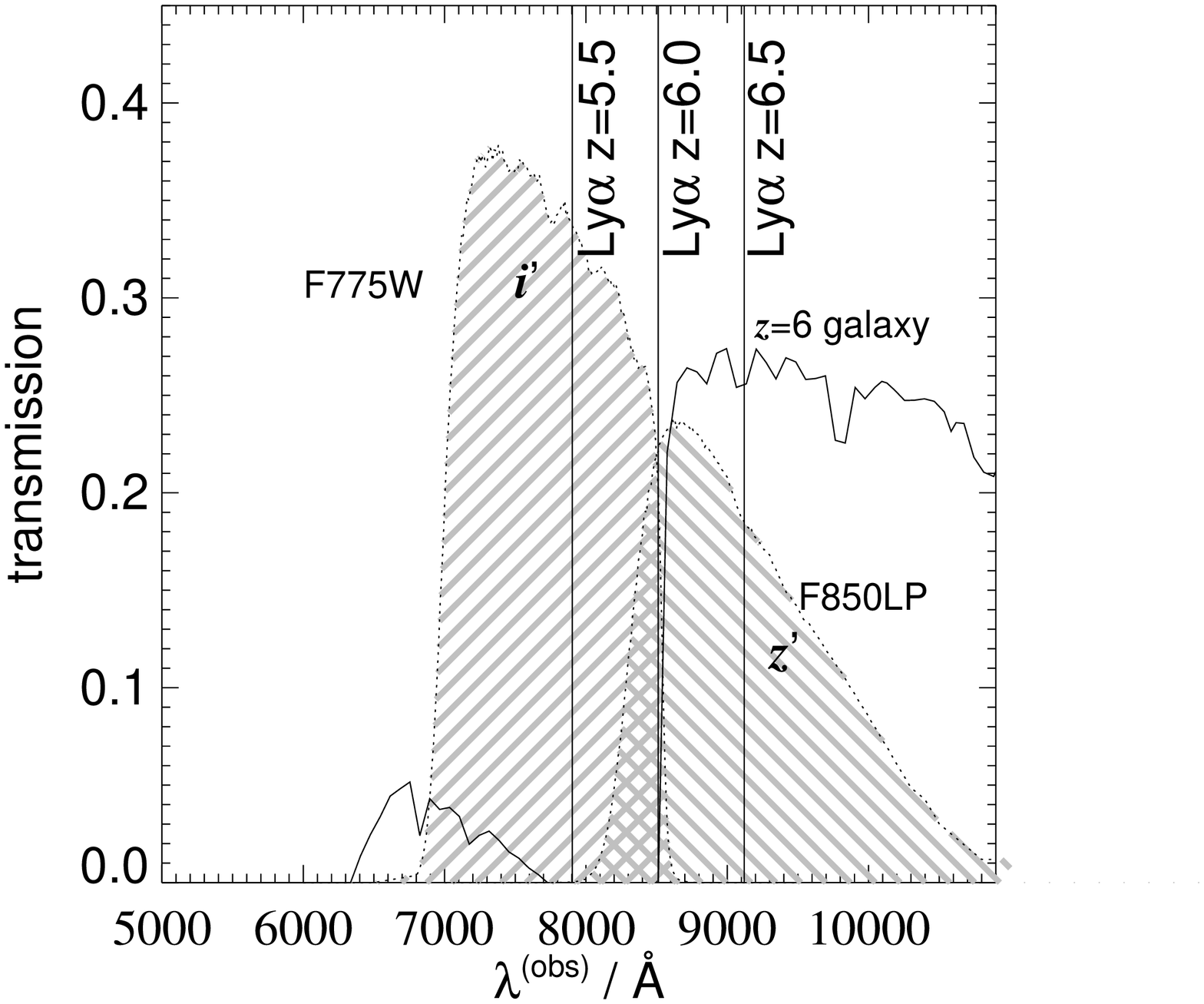}}
\caption{The ACS-$i'$ and -$z'$ bandpasses overplotted on the
spectrum of a generic $z=6$ galaxy (solid line), illustrating the
utility of our two-filter technique for locating $z\approx 6$
sources.}
\label{fig:filters}
\end{figure}

\begin{figure}
\resizebox{0.48\textwidth}{!}{\includegraphics{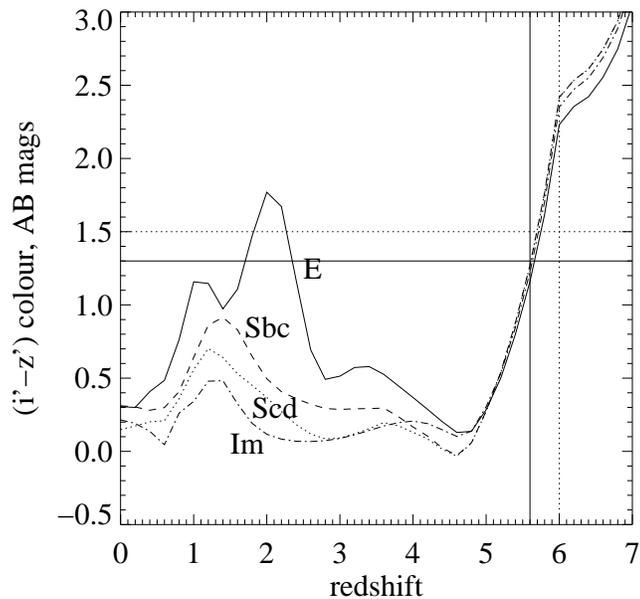}}
\caption{Model colour-redshift tracks for galaxies with non-evolving
stellar populations (from Coleman, Wu \& Weedman 1980 template
spectra). The contaminating `hump' in the $(i'-z')$ colour at
$z\approx 1-2$ arises when the Balmer break and/or the 4000\,\AA\
break redshifts beyond the $i'$-filter. }
\label{fig:tracks}
\end{figure}

\begin{figure}
\resizebox{0.48\textwidth}{!}{\includegraphics{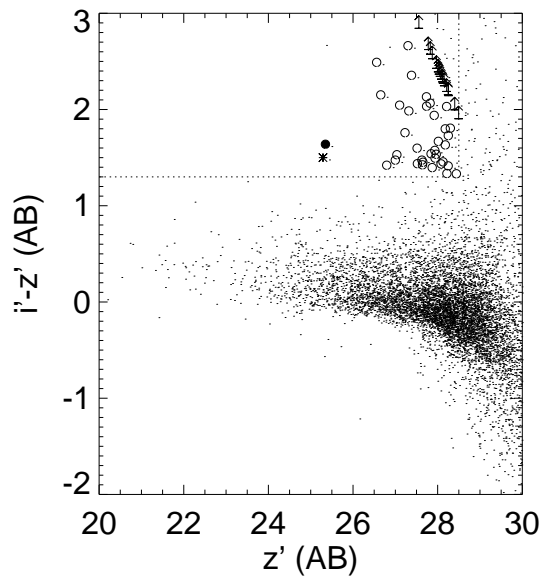}}
\caption{Colour-magnitude diagram for the UDF data with the limit
  $z'_{AB}<28.5$ and $(i'-z')_{AB}=1.3$ colour cut shown
  (dashed lines). As discussed in the text, such a catalogue could be
  contaminated by cool stars, EROs and wrongly identified extended
  objects and diffraction spikes but nonetheless provides a secure
  upper limit to the abundance of z$\approx 6$ star forming galaxies.
  Circles and arrows (lower limits) indicate our $i'$-drop candidate
  $z\approx 6$ galaxies. The solid circle is the
  spectroscopically-confirmed galaxy SBM03\#1 (Stanway et al.\ 2004b;
  Dickinson et al.\ 2004), and the asterisk is the only unresolved
$i'$-drop in our UDF sample, the probable star \#11337.}
\label{fig:izzcolmag}
\end{figure}

In order to select $z\approx 6$ galaxies, we use the Lyman break technique
pioneered at $z\sim 3$ using ground-based telescopes by Steidel and
co-workers and using {\em HST} by Madau et al.\ (1996). At $z\sim 3-4$
the technique involves the use of three filters: one below the Lyman
limit ($\rm \lambda_{rest}=912$\,\AA ), one in the Lyman forest region
and a third longward of the Lyman-$\alpha$ line ($\rm
\lambda_{rest}=1216$\,\AA). At $z\approx 6$, we can efficiently use only
two filters, since the integrated optical depth of the Lyman-$\alpha$
forest is $\gg 1$ (see Figure~\ref{fig:filters}) rendering the
shortest-wavelength filter below the Lyman limit redundant. The key
issue is to work at a sufficiently-high signal-to-noise ratio that
$i'$-band drop-outs can be safely identified through detection in a
single redder band (i.e., SDSS-$z'$). This approach has been
demonstrated to be effective by the SDSS collaboration in the
detection of $z\approx 6$ quasars using the $i'$- and $z'$-bands alone
(Fan et al.\ 2001).  The sharp sides of the SDSS filters assist in the
clean selection using the photometric redshift technique.  In
Figures~\ref{fig:tracks}\,\&\,\ref{fig:izzcolmag} we illustrate how a
colour cut of $(i'-z')_{AB}>1.5$ (used in Papers~I-III) can be
effective in selecting sources with $z>5.7$. Here we relax this cut to
$(i'-z')_{AB}>1.3$ to recover most galaxies at redshifts $z>5.6$,
but at the expense of potentially larger contamination by $z\approx 1-2$
ellipticals. Near-infrared colours from the NICMOS imaging of the UDF
should identify these Extremely Red Objects (EROs), and we consider
this in a companion paper (Stanway, McMahon \& Bunker 2004c).

\begin{figure}
\resizebox{0.98\columnwidth}{!}{\includegraphics{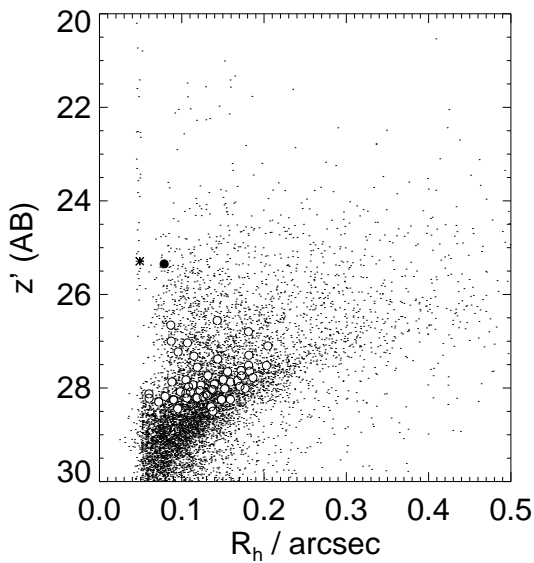}}
\caption{The distribution of angular sizes (half-light radius, $R_h$,
in arcseconds) for objects in our $z'$-band selected catalogue. Our
$i'$-drop candidate $z\approx 6$ are marked as open circles, with the
confirmed $z=5.8$ galaxy SBM03\#1 a solid circle.  The $i'$-drops
appear to be compact but resolved (the stellar locus at $0\farcs05$ is
clearly visible). The asterisk denotes the only unresolved $i'$-drop
in our UDF sample, the probable star \#11337.}
\label{fig:fwhm}
\end{figure}

Six of the 63 candidate $i'$-dropouts in our $z'_{AB}<28.5$ UDF
catalogue were identified visually as different regions of the same
extended source, and where these were within our aperture diameter of
$0\farcs5$ the duplicates were eliminated from the final
selection. One spurious $i'$-drop arose from the diffraction spikes of
bright stars due to the more extended PSF in the $z'$-band compared
with that in the $i'$-band. Only one of the $i'$-dropouts is
unresolved (Figure~\ref{fig:fwhm}). This is the brightest at
$z'_{AB}=25.3$ (\#11337 in Table~1), detected in the $V$-band image
and removed from our catalogue of potential $z\approx 6$ objects as a
probable star. At the edge of the UDF frame (and outside the central
11\,arcmin$^{2}$ region of lowest noise where we do our main analysis)
there is a second unresolved $i'$-drop with $z'_{AB}=25.2$, first
identified in Paper~I (SBM03\#5), where we argued that the near-IR
colours are likely to be stellar. It is interesting that the level of
stellar contamination in the UDF $i'$-drops is only 2\%, compared with
about one in three at the bright end ($z'_{AB}<25.6$, Papers I \&
III). This may be because we are seeing through the Galactic disk at
these faint magnitudes to a regime where there are no stars at these
faint limiting magnitudes.

\begin{figure}
\resizebox{0.98\columnwidth}{!}{\includegraphics{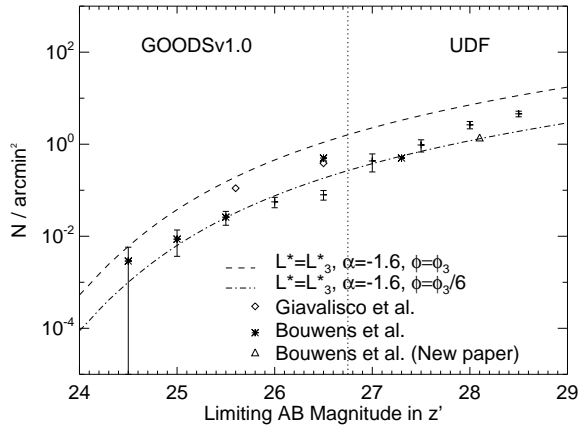}}
\caption{Cumulative source counts per arcmin$^{2}$ of $i'$-dropouts
as a function of $z'$-band magnitude. The new UDF data (over a
smaller area of 11\,arcmin$^{2}$ for $z'_{AB}\ge 27.0$) is compared
with $z'_{AB}\le 25.6$ single epoch GOODSv0.5 ACS/WFC imaging over
300\,arcmin$^{2}$ (Papers~I-III) and combined 5 epoch GOODSv1.0
images to $z'_{AB}<27.0$ (Stanway 2004).}
\label{fig:numcounts}
\end{figure}

From our original list of 63 $i'$-drops, 6 duplications were removed,
along with one diffraction spike artifact. The remaining objects
satisfying our $(i'-z')_{AB}>1.3$ \& $z'_{AB}<28.5$ selection criteria
are detailed in Table~1, of which 54 are good candidate $z\approx 6$
galaxies, along with the probable star \#11337, and another objected
(\#46574) detected in $V$-band. The surface density of $i'$-drops as a
function of limiting magnitude is shown in Figure~\ref{fig:numcounts}.
None of the $i'$-drops (with the exception of the Galactic star) are
detected in the $B$-band image of the UDF, to a $3\,\sigma$ limit of
$B_{AB}>29.2$ in a 0\farcs 5-diameter aperture, as would be expected
for the $z\approx 6$ interpretation where the $B$-band filter covers
wavelengths below the 912\,\AA\ Lyman limit.

\begin{figure}
\resizebox{0.48\textwidth}{!}{\includegraphics{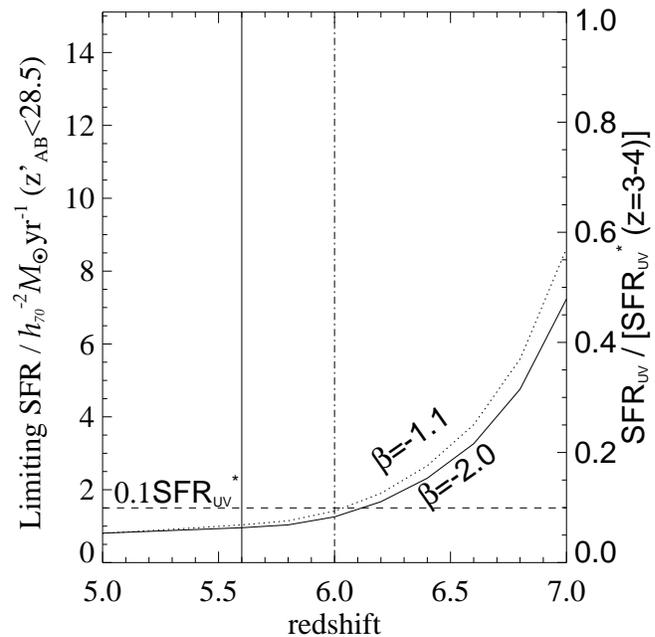}}
\caption{Limiting star formation rate as a function of redshift for
the UDF catalogue with $z'_{AB}<28.5$\,mag ($8\,\sigma$). Star
formation rates are inferred from the rest-frame 1500\,\AA\ flux
(Madau, Pozzetti \& Dickinson 1998) taking account of $k$-corrections,
filter transmission and blanketting by Lyman-$\alpha$ absorption.  The
solid line assumes a spectral slope $\beta=-2.0$ (where
$f_{\lambda}\propto \lambda^{\beta}$) appropriate for an unobscured
starburst, and the dotted line has $\beta=-1.1$ (corresponding to mean
reddening of $z\approx 3$ Lyman break galaxies given in Meurer et al.\
1997). The limit as a fraction of $L^{*}_{3}$ ($L^{*}{\rm [1500\AA ]}$
at $z\approx 3$, equivalent to ${\rm SFR}^{*}_{\rm
UV}=15\,h^{-2}_{70}\,M_{\odot}\,{\rm yr}^{-1}$ from Steidel et al.\
1999) is shown on the right axis.  Our colour selection should remove
most $z<5.6$ galaxies (solid vertical line), and our average $i'$-drop
redshift for $z'<28.5$ should be $z\approx 6.0$ (vertical dot-dash
line): we are sensitive as faint at $0.1\,L^{*}_{3}$ at this
redshift.}
\label{fig:sfrlimits}
\end{figure}

\begin{table*}
\caption{$i'$-band dropouts in the UDF. The two
stars are above the line -- all others are spatially resolved.
Our ID and the corresponding match from the UDF catalogues released
by STScI are listed. Where two close $i'$-drops
lie within our $0\farcs5$-diameter aperture, the flux only counted once in the star formation 
total -- those IDs and star formation rates in parentheses are
not counted. The star formation rates assume the $i'$-drops lie
at $z=6.0$, the expected median redshift of our sample. The $z'_{AB}$ (total) is the SExtractor ``MAG\_AUTO''.} 
\begin{tabular}{c|c|c|c|c|c|c|c|c|c}
\hline\hline
Our ID & STScI & RA \& Declination  &  $z'_{AB}$  & $i'_{AB}$ & $(i'-z')_{AB}$   & R$_h$ & $z'_{AB}$ & SFR$^{z=6}_{\rm UV}$ \\
 & ID & (J2000)  & \multicolumn{2}{c}{(0\farcs5-diameter aperture)} & 0\farcs5-aper
& arcsec & (total) & $h^{-2}_{70}\,M_{\odot}\,{\rm yr}^{-1}$ \\
\hline\hline
[(2140)$^{\star}$ & --- & 03 32 38.80  $-$27 49 53.6 &   25.22 $\pm$   0.02 &   27.91 $\pm$   0.04 
&    2.69 $\pm$   0.05 &  0.06 & 25.17 $\pm$ 0.02 & (star)] \\
 (11337) & 443 & 03 32 38.02  $-$27 49 08.4 & 25.29 $\pm$   0.02 & 26.79 $\pm$   0.04 
&  1.50 $\pm$   0.05 & 0.05 &  25.43 $\pm$ 0.02 & (star) \\
\hline
20104$^{1}$ &  2225 & 03 32 40.01  $-$27 48 15.0 &   25.35 $\pm$   0.02 & 
  26.99 $\pm$   0.03 &    1.64 $\pm$   0.04 &  0.08 & 25.29 $\pm$ 0.02 & 19.5[$z=5.83$]\\
42929$^{2}$ &  8033 & 03 32 36.46  $-$27 46 41.4 &   26.56 $\pm$   0.03 & 
  29.05 $\pm$   0.14 &    2.49 $\pm$   0.15 &  0.14 & 26.55 $\pm$ 0.04 &  6.75 \\
41628 &  8961 & 03 32 34.09  $-$27 46 47.2 &   26.65 $\pm$   0.04 & 
  28.81 $\pm$   0.12 &    2.15 $\pm$   0.12 &  0.09 & 26.70 $\pm$ 0.04 &
6.18 \\
(46574)$^{3}$ & 7730 & 03 32 38.28  $-$27 46 17.2 &   26.71 $\pm$   0.04 & 
  29.38 $\pm$  0.18 &    2.67 $\pm$   0.18 & 0.09 & 26.74 $\pm$ 0.04 &
(5.87) \\
24019 &  3398 & 03 32 32.61  $-$27 47 54.0 &   26.80 $\pm$   0.04 & 
  28.22 $\pm$   0.08 &    1.42 $\pm$   0.09 &  0.18 & 26.73 $\pm$ 0.04 &  5.42 \\
52880 & 9857 &  03 32 39.07  $-$27 45 38.8 &   27.00 $\pm$   0.05 & 
  28.47 $\pm$   0.09 &    1.47 $\pm$   0.10 &  0.09 & 27.10 $\pm$ 0.05 &  4.50 \\
23516 & 3325 &  03 32 34.55  $-$27 47 56.0 &   27.04 $\pm$   0.05 & 
  28.57 $\pm$   0.10 &    1.53 $\pm$   0.11 &  0.11 & 27.05 $\pm$ 0.05 &  4.35 \\
10188 & 322 &  03 32 41.18  $-$27 49 14.8 &   27.10 $\pm$   0.05 &  
 29.15 $\pm$   0.16 &    2.04 $\pm$   0.16 &  0.20 & 27.06 $\pm$ 0.05 &  4.10 \\
21422 & 2690 &  03 32 33.78  $-$27 48 07.6 &   27.23 $\pm$   0.05 & 
  28.99 $\pm$   0.14 &    1.76 $\pm$   0.15 &  0.10 & 27.37 $\pm$ 0.05 &  3.64 \\
25578$^{D}$ & --- &  03 32 47.85  $-$27 47 46.4 &   27.30 $\pm$   0.06 & 
  29.96 $\pm$   0.31 &    2.66 $\pm$   0.31 &  0.18 & 27.28 $\pm$ 0.06 &  3.41 \\
25941 & 4050 &  03 32 33.43  $-$27 47 44.9 &   27.32 $\pm$   0.06 & 
  29.30 $\pm$   0.18 &    1.99 $\pm$   0.19 &  0.11 & 27.38 $\pm$ 0.06 &  3.35 \\
26091$^{D}$ & 4110 &  03 32 41.57  $-$27 47 44.2 &   27.38 $\pm$   0.06 &  
 29.74 $\pm$   0.25 &    2.35 $\pm$   0.26 &  0.14 & 27.21 $\pm$ 0.07 &  3.16 \\
24458 & 3630 &  03 32 38.28  $-$27 47 51.3 &   27.51 $\pm$   0.07 &  
 29.11 $\pm$   0.15 &    1.60 $\pm$   0.17 &  0.18 & 27.67 $\pm$ 0.08 &  2.80 \\
21262 & 2624 &  03 32 31.30  $-$27 48 08.3 &   27.52 $\pm$   0.07 &  
 28.96 $\pm$   0.13 &    1.44 $\pm$   0.15 &  0.20 & 27.49 $\pm$ 0.08 &  2.78 \\
13494 & 30591 &  03 32 37.28  $-$27 48 54.6 &   27.56 $\pm$   0.07 &  
 30.62 $\pm$   0.55 &    3.06 $\pm$   0.55 &  0.12 & 27.48 $\pm$ 0.08 &  2.69 \\
24228 & 3450 &  03 32 34.28  $-$27 47 52.3 &   27.63 $\pm$   0.07 &  
 29.10 $\pm$   0.15 &    1.47 $\pm$   0.17 &  0.17 & 27.39 $\pm$ 0.08 &  2.52 \\
16258 &  1400 & 03 32 36.45  $-$27 48 34.3 &   27.64 $\pm$   0.07 & 
  29.07 $\pm$   0.15 &    1.42 $\pm$   0.16 &  0.18 & 27.25 $\pm$ 0.07 &  2.49 \\
42414 & 9202 &  03 32 33.21  $-$27 46 43.3 &   27.65 $\pm$   0.07 &  
 29.10 $\pm$   0.15 &    1.45 $\pm$   0.17 &  0.16 & 27.54 $\pm$ 0.08 &  2.46 \\
27173$^{5}$ & 4377 &  03 32 29.46  $-$27 47 40.4 &   27.73 $\pm$   0.08 
&   29.87 $\pm$   0.28 &    2.13 $\pm$   0.29 &  0.13 & 27.74 $\pm$ 0.09 &  2.28 \\
49117$^{D}$ & --- &  03 32 38.96  $-$27 46 00.5 &   27.74 $\pm$   0.08 
&   29.77 $\pm$   0.26 &    2.03 $\pm$   0.27 &  0.17 & 27.36 $\pm$ 0.07 &  2.28 \\
49701 & 36749 &  03 32 36.97  $-$27 45 57.6 &   27.78 $\pm$   0.08 
&   30.79 $\pm$   0.64 &    3.02 $\pm$   0.64 &  0.19 &  27.90 $\pm$ 0.09 &  2.20 \\
24123 & --- &  03 32 34.29  $-$27 47 52.8 &   27.82 $\pm$   0.08 
&   29.89 $\pm$   0.29 &    2.07 $\pm$   0.30 &  0.15 &  27.65 $\pm$ 0.09 &  2.11 \\
27270 & 33003 &  03 32 35.06  $-$27 47 40.2 &   27.83 $\pm$   0.08 
&   30.69 $\pm$   0.58 &    2.87 $\pm$   0.59 &  0.11 &  27.99 $\pm$ 0.09 &  2.10 \\
23972 &  3503 & 03 32 34.30  $-$27 47 53.6 &   27.84 $\pm$   0.09 
&   29.38 $\pm$   0.19 &    1.54 $\pm$   0.21 &  0.17 & 27.77 $\pm$ 0.10 &  2.07 \\
14751 & 1086 &  03 32 40.91  $-$27 48 44.7 &   27.87 $\pm$   0.09 
&   29.27 $\pm$   0.17 &    1.40 $\pm$   0.19 &  0.09 & 27.92 $\pm$ 0.09 &  2.02 \\
44154 & 35945 &  03 32 37.46  $-$27 46 32.8 &   27.87 $\pm$   0.09 
&  $>$\,30.4 (3\,$\sigma$) &   $>$\,2.5 (3\,$\sigma$) &  0.16 & 27.87 $\pm$ 0.10 &  2.01 \\
35084 & 34321 &  03 32 44.70  $-$27 47 11.6 &   27.92 $\pm$   0.09 
&   29.86 $\pm$   0.28 &    1.94 $\pm$   0.30 &  0.14 & 27.90 $\pm$ 0.09 &  1.93 \\
42205 & 8904 &  03 32 33.55  $-$27 46 44.1 &   27.93 $\pm$   0.09 
&   29.51 $\pm$   0.21 &    1.57 $\pm$   0.23 &  0.11 & 27.91 $\pm$ 0.09 &  1.90 \\
46503 & 7814 &  03 32 38.55  $-$27 46 17.5 &   27.94 $\pm$   0.09 
&   29.43 $\pm$   0.20 &    1.50 $\pm$   0.22 &  0.12 & 28.07 $\pm$ 0.09 &  1.89 \\
19953 & 2225 &  03 32 40.04  $-$27 48 14.6 &   27.97 $\pm$   0.09 
&   29.50 $\pm$   0.21 &    1.54 $\pm$   0.23 &  0.17 & 27.68 $\pm$ 0.10 &  1.85 \\
52086 & 36786 &  03 32 39.45  $-$27 45 43.4 &   27.97 $\pm$   0.09 
&   30.83 $\pm$   0.66 &    2.86 $\pm$   0.66 &  0.11 & 28.04 $\pm$ 0.10 &  1.84 \\
44194 & 35945 &  03 32 37.48  $-$27 46 32.5 &   28.01 $\pm$   0.10 
&   30.61 $\pm$   0.54 &    2.60 $\pm$   0.55 &  0.18 & 27.46 $\pm$ 0.09 &  1.77 \\
21111$^{D}$ & 2631 &  03 32 42.60  $-$27 48 08.9 &   28.02 $\pm$   0.10
 &   29.69 $\pm$   0.24 &    1.67 $\pm$   0.26 &  0.15 & 28.08 $\pm$ 0.10 &  1.76 \\
46223$^{4}$ & 35506 &  03 32 39.87  $-$27 46 19.1 &   28.03 $\pm$   0.10 
&   32.18 $\pm$   2.23 &    4.15 $\pm$   2.23 &  0.14 & 28.10 $\pm$ 0.11 &  1.74 \\
22138 & 32007 & 03 32 42.80  $-$27 48 03.2 &   28.03 $\pm$   0.10 
&   $>$\,30.4 (3\,$\sigma$) &  $>$\,2.3 (3\,$\sigma$) &  0.14 & 28.14 $\pm$ 0.10 &  1.73 \\
(46234)$^{4}$ & --- &  03 32 39.86  $-$27 46 19.1 &   28.05 $\pm$   0.10 
&   30.61 $\pm$   0.54 &    2.56 $\pm$   0.55 &  0.12 & 28.30 $\pm$ 0.12 &  (1.70) \\
14210 & 978 &  03 32 35.82  $-$27 48 48.9 &   28.08 $\pm$   0.10 
&   29.51 $\pm$   0.21 &    1.43 $\pm$   0.24 &  0.10 & 28.16 $\pm$ 0.11 &  1.66 \\
45467 & 35596 & 03 32 43.02  $-$27 46 23.7 &   28.08 $\pm$   0.10 
&    $>$\,30.4 (3\,$\sigma$) &   $>$\,2.3 (3\,$\sigma$) &  0.11 & 28.25 $\pm$ 0.10 &  1.66 \\
12988$^{D}$ & 30534 & 03 32 38.49  $-$27 48 57.8 &   28.11 $\pm$   0.11 
&   30.47 $\pm$   0.48 &    2.36 $\pm$   0.49 &  0.10 & 28.22 $\pm$ 0.11 &  1.61 \\
30359 & 33527 & 03 32 30.14  $-$27 47 28.4 &   28.13 $\pm$   0.11 
&   29.58 $\pm$   0.22 &    1.46 $\pm$   0.25 &  0.13 & 28.02 $\pm$ 0.11 &  1.59 \\
11370 & 482 & 03 32 40.06  $-$27 49 07.5 &   28.13 $\pm$   0.11 
&   30.45 $\pm$   0.47 &    2.32 $\pm$   0.48 &  0.06 & 28.27 $\pm$ 0.08 &  1.59 \\
24733 & 32521 & 03 32 36.62  $-$27 47 50.0 &   28.15 $\pm$   0.11 
&   30.92 $\pm$   0.71 &    2.76 $\pm$   0.72 &  0.13 & 28.34 $\pm$ 0.12 &  1.55 \\
37612 & 34715 & 03 32 32.36  $-$27 47 02.8 &   28.18 $\pm$   0.11 
&   29.98 $\pm$   0.31 &    1.80 $\pm$   0.33 &  0.13 & 28.15 $\pm$ 0.11 &  1.52 \\
41918 & 7829 & 03 32 44.70  $-$27 46 45.5 &   28.18 $\pm$   0.11 
&   29.81 $\pm$   0.27 &    1.63 $\pm$   0.29 &  0.08 & 28.36 $\pm$ 0.10 &  1.52 \\
21530 & 31874 & 03 32 35.08  $-$27 48 06.8 &   28.21 $\pm$   0.12 
&   30.24 $\pm$   0.39 &    2.03 $\pm$   0.41 &  0.12 & 28.35 $\pm$ 0.12 &  1.47 \\
42806 & 8033 & 03 32 36.49  $-$27 46 41.4 &   28.21 $\pm$   0.12 
&   30.76 $\pm$   0.62 &    2.55 $\pm$   0.63 &  0.11 & 28.12 $\pm$ 0.11 &  1.47 \\
27032$^{5}$ & 4377 &  03 32 29.45  $-$27 47 40.6 &   28.22 $\pm$   0.12 
&   29.55 $\pm$   0.22 &    1.34 $\pm$   0.25 &  0.06 & 28.70 $\pm$ 0.12 &  1.46 \\
52891 & 36697 & 03 32 37.23  $-$27 45 38.4 &   28.25 $\pm$   0.12 
&   32.21 $\pm$   2.28 &    3.96 $\pm$   2.28 &  0.16 & 28.34 $\pm$ 0.11 &  1.43 \\
17908 & 1834 & 03 32 34.00  $-$27 48 25.0 &   28.25 $\pm$   0.12 
&   29.66 $\pm$   0.24 &    1.41 $\pm$   0.27 &  0.15 & 28.22 $\pm$ 0.13 &  1.42 \\
(27029)$^{5}$ & 4353 & 03 32 29.44  $-$27 47 40.7 &   28.25 $\pm$   0.12 
&   29.98 $\pm$   0.31 &    1.73 $\pm$   0.33 &  0.09 & 28.67 $\pm$ 0.14 &  (1.42) \\
48989$^{D}$ & 36570 & 03 32 41.43  $-$27 46 01.2 &   28.26 $\pm$   0.12 
&    $>$\,30.4 (3\,$\sigma$) &   $>$\,2.1  (3\,$\sigma$)&  0.09 & 28.45 $\pm$ 0.12 &  1.41 \\
17487 & --- & 03 32 44.14  $-$27 48 27.1 &   28.30 $\pm$   0.12 
&   30.10 $\pm$   0.35 &    1.81 $\pm$   0.37 &  0.07 & 28.51 $\pm$ 0.11 &  1.36 \\
18001 & 31309 & 03 32 34.14  $-$27 48 24.4 &   28.40 $\pm$   0.13 
&   30.46 $\pm$   0.48 &    2.06 $\pm$   0.49 &  0.14 & 28.59 $\pm$ 0.14 &  1.23 \\
35271 & 6325 & 03 32 38.79  $-$27 47 10.9 &   28.44 $\pm$   0.14
&   29.77 $\pm$   0.26 &    1.33 $\pm$   0.30 &  0.10 & 28.60 $\pm$ 0.13 &  1.19 \\
22832 & --- & 03 32 39.40  $-$27 47 59.4 &   28.50 $\pm$   0.15 
&   30.46 $\pm$   0.47 &    1.96 $\pm$   0.50 &  0.14 & 28.60 $\pm$ 0.13 &  1.13 \\
\hline
\end{tabular}
\label{tab}
\ \\
$^{D}$\,double.\
$^{\star}$\,star SBM03\# 5 (Paper~I), outside central UDF.\
$^{1}$\,SBM03\#1 (Paper~I); SiD002 (Dickinson et al.\ 2004).\
$^{2}$\,SiD025 (Dickinson et al.\ 2004).\
$^{3}$\,46574 has a close neighbour visible in the $v$-band (i.e.\ low
redshift.)\
$^{4}$\,46234 is close to 46223.\
$^{5}$\,27029 is close to 27032.
\end{table*}


\section{Selection Effects and The Luminosity Function of Star Forming 
Galaxies at $z\sim 6$}
\label{sec:discuss}

\subsection{Estimate of Star Formation Rate from the Rest-UV}

We will base our measurement of the star formation rate for each
candidate on the rest-frame UV continuum, redshifted into the
$z'$-band at $z\approx 6$ and measured from the counts in a
$0\farcs5$-diameter aperture (with an aperture correction to total
magnitudes, Section~\ref{sec:cats}).  In the absence of dust
obscuration, the relation between the flux density in the rest-UV
around $\approx 1500$\,\AA\ and the star formation rate (${\rm SFR}$
in $M_{\odot}\,{\rm yr}^{-1}$) is given by $L_{\rm UV}=8\times 10^{27}
{\rm SFR}\,{\rm ergs\,s^{-1}\,Hz^{-1}}$ from Madau, Pozzetti \&
Dickinson (1998) for a Salpeter (1955) stellar initial mass function
(IMF) with $0.1\,M_{\odot}<M^{*}<125\,M_{\odot}$. This is comparable
to the relation derived from the models of Leitherer \& Heckman (1995)
and Kennicutt (1998).  However, if a Scalo (1986) IMF is used, the
inferred star formation rates will be a factor of $\approx 2.5$ higher
for a similar mass range.

Recognising the limitations of our earlier studies (Papers~I-III)
which by necessity focussed on the brighter $i'$-drops, we now attempt
to recover the $z\approx 6$ rest-frame UV luminosity function from the
observed number counts of $i'$-drops to faint magnitudes in the UDF.
Although our colour cut selects galaxies with redshifts in the range
$5.6<z<7.0$, an increasing fraction of the $z'$-band flux is
attenuated by the redshifted Lyman-$\alpha$ forest. At higher
redshifts we probe increasingly shortward of $\lambda_{\rm
rest}=1500$\,\AA\ (where the luminosity function is calculated) so the
$k$-corrections become significant beyond $z\approx 6.5$.

Figure~\ref{fig:sfrlimits} demonstrates this bias and shows the
limiting star formation rate as a function of redshift calculated by
accounting for the filter transmissions and the blanketting effect of
the intervening Lyman-$\alpha$ forest. By introducing the small
$k$-correction to $\lambda_{\rm rest}=1500$\,\AA\ from the observed
rest-wavelengths longward of Lyman-$\alpha$ redshifted into the
$z'$-band we can correct for this effect. We considered a spectral
slope of $\beta=-2.0$ (where $f_{\lambda}\propto \lambda^{\beta}$)
appropriate for an unobscured starburst (flat in $f_{\nu}$), and also
a redder slope of $\beta=-1.1$ which appropriate for mean reddening of
the $z\approx 3$ Lyman break galaxies given by Meurer et al.\
(1997). A more recent determination for this population by Adelberger
\& Steidel (2000) gives $\beta=-1.5$, in the middle of the range. At our
$8\,\sigma$ limiting magnitude of $z'_{AB}=28.5$, we deduce we can
detect unobscured star formation rates as low as
$1.0\,[1.1]\,h^{-2}_{70}\,M_{\odot}\,{\rm yr}^{-1}$ at $5.6<z<5.8$ and
$1.5\,[1.7]\,h^{-2}_{70}\,M_{\odot}\,{\rm yr}^{-1}$ at $z<6.1$ for
spectral slope $\beta=-2.0\,[-1.1]$ (Figure~\ref{fig:sfrlimits}).

Recognising that contamination by interlopers will only reduce the
value, we now compare the comoving star formation rate deduced for
$z\approx 6$ galaxies based on our candidate $i'$-dropout source counts
with predictions based on a range of rest-frame UV luminosity
functions. For convenience we assume that there is no evolution over the
sampled redshift range, $5.6<z<6.5$, spanned by the UDF data (equivalent
to a range between $0.8-1.0\,h_{70}^{-1}$\,Gyr after the Big Bang). We
take as a starting point the luminosity function for the well-studied
Lyman-break $U$-dropout population, reported in Steidel et al.\ (1999),
which has a characteristic rest-UV luminosity $m^{*}_{R}=24.48$
(equivalent to $M^{*}_{3}(1500\,{\rm \AA})=-21.1$\,mag or
$L^{*}_{3}=15\,h^{-2}_{70}\,M_{\odot}\,{\rm yr}^{-1}$ for our
cosmology). The faint end slope of the Schechter function at $z\approx
3$ is relatively steep ($\alpha=-1.6$) compared with $\alpha=-1.0$ to
$-1.3$ for lower-redshift galaxy samples (e.g., Lilly et al.\ 1995;
Efstathiou et al.\ 1988; Blanton et al.\ 2003 -- see Gabasch et al.\
2004 for recent determinations at 1500\,\AA ). The characteristic
comoving number density at $z\approx 3$ is
$\Phi^{*}_{3}=0.00138\,h^{3}_{70}\,{\rm Mpc}^{-3}\,{\rm mag}^{-1}$ in
our cosmology.

We adopt two approaches to determining the galaxy number density and
star formation density at $z\approx 6$: the first method
(Section\,\ref{sec:EffectiveVolume}) is the one used in Papers
I\,\&\,III, an application of the ``effective volume'' technique
(Steidel et al.\ 1999). The second method
(Section\,\ref{sec:simulations_numdens}) involved comparing the
measured surface density of $i'$-dropout $z\approx 6$ galaxies with
that predicted on the assumption they have the same characteristics as
the $U$-dropout population at $z\approx 3$.

\subsubsection{Effective Survey Volume}
\label{sec:EffectiveVolume}

We have followed the approach of Steidel et al.\ in calculating the
effect of luminosity bias on our sample of $z\approx 6$ LBGs.  We
account for the $k$-correction: as redshift increases, the $z'$-band
samples light in the rest-frame of the galaxies at wavelengths that
are increasingly far to the blue of 1500\,\AA , where the LBGs'
luminosity function was calculated.  Additionally, at redshifts $z>6$,
Lyman-$\alpha$ absorption from the forest enters the $z'$-band and
makes galaxies fainter still, as there is incomplete coverage of the
filter by the continuum longward of Lyman-$\alpha$.  Accounting for
these luminosity and redshift biases, we compute an effective survey
volume using
\[
V_{\rm eff}(m)=\int dz\,p(m,z)\,\frac{dV}{dz}
\]
where $p(m,z)$ is the probability of detecting a galaxy at redshift $z$
and apparent $z'$ magnitude $m$, and $dz\,\frac{dV}{dz}$ is the comoving
volume per unit solid angle in a slice $dz$ at redshift $z$. We
integrate over the magnitude range we are sensitive to, and over the
redshift range $5.6<z<7.0$ from our colour selection, and calculate that
for a spectral slope of $\beta=-2.0$ (i.e., flat in $f_{\nu}$) the
effective comoving volume is 40 per cent the total volume in the range
$5.6<z<7.0$ (i.e., the same as $5.6<z<6.1$). For our 11\,arcmin$^{2}$
survey area (excluding the edge regions of the UDF where fewer
frames overlap) this is a comoving volume of $2.6\times
10^{4}\,h^{-3}_{70}\,{\rm Mpc}^{3}$. 

\begin{figure}
\resizebox{0.98\columnwidth}{!}{\includegraphics{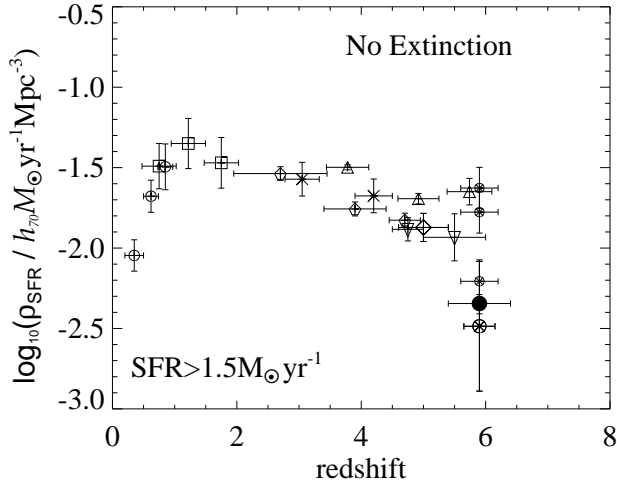}}
\caption{An updated version of the `Madau-Lilly' diagram (Madau et
al.\ 1996; Lilly et al.\ 1996) illustrating the evolution of the
comoving volume-averaged star formation rate.  Our work from the UDF
data is plotted a solid symbol. Other determinations have been
recalculated for our cosmology and limiting UV luminosity of
$1.5\,h_{70}^{-2}\,M_{\odot}\,{\rm yr}^{-1}$ at $z=6.1$ (equivalent to
$0.1\,L^*_3$ at $z\approx 3$ from Steidel et al.\ 1999), assuming a
slope of $\alpha=-1.6$ for $z>2$ and $\alpha=-1.3$ for $z<2$. Data
from the CFRS survey of Lilly et al.\ (1996) are shown as open
circles; data from Connolly et al.\ (1997) are squares; and the Lyman
break galaxy work of Steidel et al.\ (1999) is plotted as crosses, of
Fontana et al.\ (2002) as inverted triangles and that by Iwata et al.\
(2003) as an open diamond. Pentagons are from Bouwens, Broadhurst \&
Illingworth (2003). The upright triangles are the GOODS
$i'$-drop results from Giavalisco et al.\ (2004). The three ACS
estimates of Bouwens et al.\ (2003) are shown by small crossed circles
and indicate three different completeness corrections for one sample
of objects -- the larger symbol is the recent re-determination using a
new catalogue by this group from a deeper dataset (the UDF flanking
fields -- Bouwens et al.\ 2004); we have recomputed the comoving
number density from the Bouwens et al.\ (2004) because of a
discrepancy on the scale of their plot of star formation history
(their Fig.~4 in astro-ph/0403167\,v1\,\&\,v2).}
\label{fig:madau_plot}
\end{figure}

Hence we calculate a volume-averaged comoving star formation density at
$z\approx 6$ of $(0.005\pm 0.001)\,h_{70}\,M_{\odot}\,{\rm
yr}^{-1}\,{\rm Mpc}^{-3}$ for the $\approx 50$ $i'$-dropout galaxies
with $z'(AB)<28.5$ ($L_{UV}>0.1\,L^{*}_{3}$). This is plotted on the
Madau-Lilly diagram (Figure~\ref{fig:madau_plot}). Data from other
groups are shown on this figure, where we have corrected all the
datasets to the same limiting star formation rate of
$1.5\,h_{70}^{-2}\,M_{\odot}\,{\rm yr}^{-1}$ (i.e., typically
integrating their claimed luminosity functions down to $\approx
0.1\,L*_{3}$) to provide a fair comparison of evolution. Integrating the
luminosity function down to $\approx 0.1\,L*$, as here, represents most
of the total luminosity density for faint end slopes $\alpha> -1.6$
(compared with integrating to zero luminosity). If we assume that
the Schechter function holds for the unobserved faint galaxies with
$L<0.1\,L*_{3}$, then the
observed population with $L>0.1\,L^{*}_{3}$ represents (87.5\%, 78.9\%,
56.4\%, 32.4\%, 17.4\%) of the total star formation rate for faint-end slopes
$\alpha=(-1.1, -1.3, -1.6, -1.8, -1.9)$.

\subsubsection{Colour Selection: spectral slope and forest blanketting}
\label{sec:colour_selection}

We model the effect of the break below the
Lyman-$\alpha$ emission line due to blanketting by the forest,
where the continuum break $D_A$ (Oke \& Korycansky 1982) is
defined as
\begin{equation}
D_{A}=\left( 1 - \frac{f_{\nu}(1050-1170\,{\rm \AA})_{\rm
obs}}{f_{\nu}(1050-1170\,{\rm \AA})_{\rm pred}}\right) .
\end{equation}
We assumed flux decrements of $D_A=0.9-1.0$, consistent with that
observed in the $z>5.8$ SDSS QSOs (Fan et al.\ 2001). We find
that lowering $D_A$ reduces the completeness in the lowest redshift bin
$5.6<z<5.8$ for a $(i'-z')_{AB}>1.5$ colour cut. A
$(i'-z')_{AB}>1.3$ cut improves the selection somewhat but at the
risk of higher contamination from red objects at $z\approx 1-2$:
we consider this in Stanway, McMahon \& Bunker (2004c).

We find that altering the spectral slope $\beta$ of the $i'$-drop
spectral energy distribution ($f_{\lambda}\propto \lambda^{-\beta}$)
over the range $-1.1>\beta>-2.0$ changes the predicted number of
$i'$-dropouts by only $\approx 10$\%.

\subsubsection{Surface Density Predictions}
\label{sec:simulations_numdens}

First, we compare our observed number of $i'$-band dropout galaxies
with a simple no-evolution model, which assumes the same luminosity
function for Lyman break galaxies at $z=6$ as at $z=3$ (with a
faint-end slope $\alpha=-1.6$, spectral slope $\beta=-2.0$ and
Lyman-$\alpha$ forest decrement $D_{A}=1.0$). This no-evolution model
would predict 169 galaxies satisfying our $i'_{AB}<28.5$ \&
$(i'-z')_{AB}>1.3$ selection with a total star formation rate of
$866\,h^{-2}_{70}\,M_{\odot}\,{\rm yr}^{-1}$. This compares with our
observed number of 54 $i'$-drops (1/3rd that predicted), which have
a total star formation rate of $140\,h^{-2}_{70}\,M_{\odot}\,{\rm
  yr}^{-1}$ (1/6th of the no-evolution prediction). The predicted
median redshift of our $i'$-drop sample for the no-evolution model is
$z=5.95$, with the luminosity-weighted average $\bar{z}=6.05$.

\begin{figure}
\resizebox{0.98\columnwidth}{!}{\includegraphics{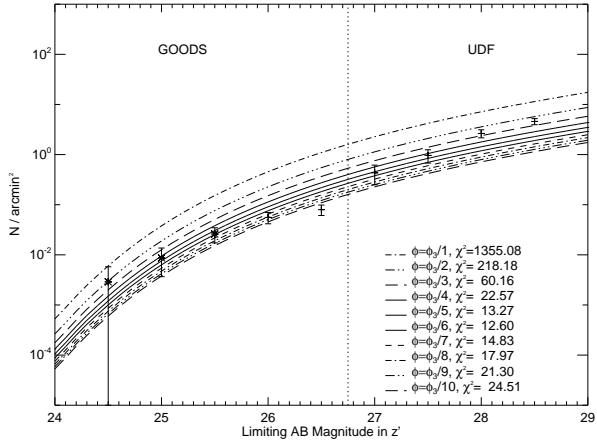}}
\caption{Cumulative source counts per arcmin$^{2}$ of $i'$-dropouts as a
function of $z'$-band magnitude, with various normalisations of the
characteristic number density at $z\approx 6$, $\Phi^{*}_{6}$ (in terms
of the value at $z\approx 3$, $\Phi^{*}_{3}$), assuming
$L^{*}_{6}=L^{*}_{3}$ and the same $\alpha$ as the $z\approx 3$ Lyman
break population ($\alpha=-1.6$). Our faintest point from the GOODS data
(at $z'_{AB}=26.5$) is excluded from the fit due to incompleteness.}
\label{fig:phi}
\end{figure}

\begin{figure}
\resizebox{0.98\columnwidth}{!}{\includegraphics{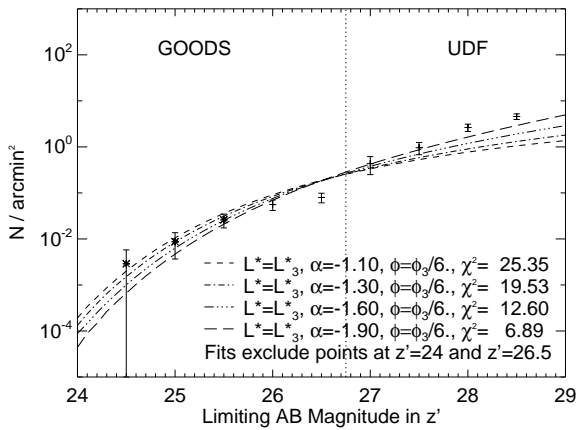}}
\caption{Cumulative source counts per arcmin$^{2}$ of $i'$-dropout
  as a function of $z'$-band magnitude, with various values of the
  faint end slope ($\alpha$) assuming $L^{*}_{6}=L^{*}_{3}$ and
  $\Phi^{*}_{6}=\Phi^{*}_{3}/6$. Our faintest point from the GOODS data (at
$z'_{AB}=26.5$) is excluded from the fit due to incompleteness.}
\label{fig:alpha}
\end{figure}

Clearly, evolution in the UV luminosity function of Lyman break
galaxies is required. To fit this, we constructed a grid of models
based upon the $z\approx 3$ luminosity function, varying $\alpha$
between $-1.1$ and $-1.9$, and $L^{*}$ between $0.3\,L^{*}_{3}$ and
$2\,L^{*}_{3}$. We leave the normalization of the luminosity function,
$\Phi^{*}$, as a free parameter,

We minimize $\chi^{2}$ for our grid of model luminosity
functions: our best fit (Figure~\ref{fig:phi}) is
compatible with no evolution of $L^*$ from $z\approx 3$,
but a large decline in the comoving space density, $\Phi^*$
(by about  a factor 6 relative to $z\approx 3$). The
faint end slope is less well constrained, although no evolution
is compatible with the results. At the faintest magnitude bin,
there modestly higher counts, perhaps indicating a slightly
steeper $\alpha$ if the results at the faintest magnitudes are
to be trusted (Figure~\ref{fig:alpha}).

\subsubsection{Comparison with $i'$-drop number counts from other groups}
\label{sec:compare_with_others}

Recognizing the very limited area of the UDF and the problems of
cosmic variance, it is nonetheless interesting to compare our measured
$i'$-drop number counts with previous determinations from shallower
data sets. The surface density derived in Paper~I to $z'_{AB}=25.6$ is
consistent with the present data -- we detect only one resolved
$i'$-dropout this bright: SBM03\#1. Note that the UDF pointing was
selected to include this object. No other spatially-resolved
$i'$-dropouts are detected to $z'_{AB}<26.5$, implying a surface
density of $0.1\pm 0.1$\,arcmin$^{-2}$. This is in contrast with the
density of $0.4$\,arcmin$^{-2}$ measured by Giavalisco et al.\ (2004)
to the same  $z'_{AB}<26.5$ limit,
and $0.5\pm 0.2$\,arcmin$^{-2}$ from the completeness-corrected
estimate of Bouwens et al.\ (2003)\footnote{{\em Note added in proof:}
a recent paper by Bouwens et al.\ (2004), based on number counts of
$i'$-drops in the ACS parallel observations to the NICMOS UDF field,
significantly revises their previous estimate of the number density of
$z'_{AB}<26.5$ $i'$-drops from $0.5\pm 0.2$ to
$0.2\pm0.1$\,arcmin$^{-2}$ (4 objects in 21\,arcmin$^{2}$,
consistent with our UDF work presented here). The conclusion of Bouwens
et al.\ (2003) --that the comoving star formation at $z\approx 6$ is
consistent with no evolution from $z\approx 4$-- is revised in Bouwens
et al.\ (2004) to be a factor of 3 decline from $z=3.8$ to $z\approx
6$. Using the evolution in comoving number density of $(1+z)^{-2.8}$
suggested by Bouwens et al.\ (2004), this fall in star formation rate
at $z=6$ is consistent with our result of a factor of 6 decline from
$z=3.0$ to $z\approx 6$ from the GOODS data in Stanway, Bunker \&
McMahon (2003), confirmed in this paper from the deeper UDF data.
}, and the even higher surface density of 2.3\,arcmin$^{-2}$ (after
removing stellar contamination) claimed by
Yan, Windhorst \& Cohen (2003), after correcting for a factor of 4 error
in their original flux calibration (see Yan \& Windhorst 2004). Clearly,
there are large discrepancies from the various groups in the number
density measured to the same limiting magnitude of $z'_{AB}<26.5$, with
measurements up to a factor of 20 higher than our UDF measurement (Yan
\& Windhorst 2004). These discrepancies may be due to cosmic variance,
or too many spurious sources in the samples of these teams, due to
working close to the sensitivity limits.  By using a high
signal-to-noise ($S/N=8$) cut, we guard against the low-$S/N$ bias:
where there are many more objects with intrinsically bluer colours that
scatter up into our $(i'-z')_{AB}>1.3$ selection than there are real
$i$-drops which scatter out of the colour selection through photometric
errors.

\begin{figure}
\resizebox{0.98\columnwidth}{!}{\includegraphics{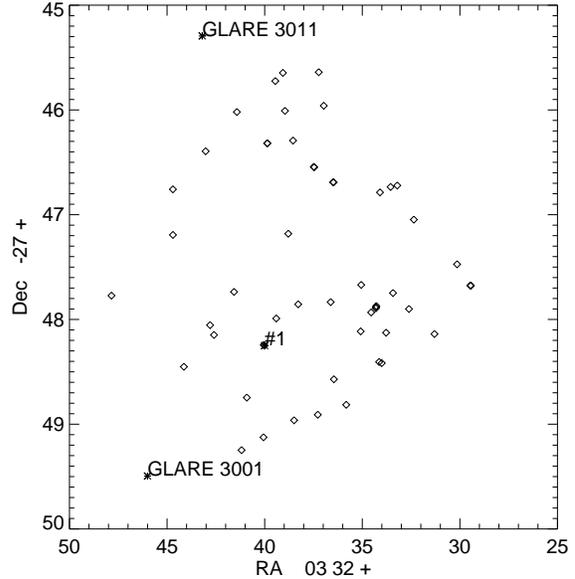}}
\caption{The spatial distribution of our UDF $i'$-drops on
the sky (diamonds). The location of the confirmed $z=5.8$ source
from Paper~I is marked (\#1) as are two other sources
just outside the UDF, spectroscopically identified at $z=5.8-5.9$
by Stanway et al.\ (2004a).}
\label{fig:distribution}
\end{figure}

From Somerville et al.\ (2004) we estimate that the cosmic variance for
the UDF is 40\%, assuming the $z=6$ LBGs are clustered in the same way
as the $z=3$ LBGs and assuming a volume of derived by scaling our UDF
area with our wider-area GOODS data (with an effective volume of
$1.8\times 10^5\,h^{-3}_{70}\,{\rm Mpc}^{3}$ for the 146\,arcmin$^{2}$
of GOODS-S, Paper~I).  Indeed, the spatial distribution of our
$i'$-drops on the sky does indicate some clustering
(Figure~\ref{fig:distribution}), and we had already flagged 6 of our
candidates as being ``double'' sources (Table~1), with another 2 having
near neighbours.  In the GLARE GMOS/Gemini spectroscopy of the
GOODS-South $i'$-dropouts, Stanway et al.\ (2004a) have already
spectroscopically identified an overdensity at $z=5.8$.

\subsection{Implications for Reionization}
\label{sec:reinonization}

The increased depth of the UDF enables us to resolve the uncertainties
associated with the unobserved portion of the luminosity function (LF)
for $z\approx 6$ sources. Our best-fit LF suggest little or no change
in $L^*$ over $3<z<6$, with $\alpha$ less well constrained but
consistent with modest evolution, implying the major evolution is a
decline in space density (and global star formation rate) by
$\simeq\times 6$ at $z\approx 6$.  This sharp decline, which must
represent a lower limit to the true decline given the likelihood of
contamination from foreground sources, suggests it may be difficult
for luminous star-forming $z\approx 6$ $i$-dropout galaxies to be the
main source of ionizing photons of the Universe.

We attempt to quantify this by comparing with the estimate
of Madau, Haardt \& Rees (1999) for the density of star formation
required for reionization (their equation 27):
\begin{equation}
{\dot{\rho}}_{\rm SFR}\approx \frac{0.013\,M_{\odot}\,{\rm yr}^{-1}\,{\rm Mpc}^{-3}}{f_{\rm esc}}\,\left( \frac{1+z}{6}\right) ^{3}\,\left( \frac{\Omega_{b}\,h^2_{50}}{0.08}\right) ^{2}\,\left( \frac{C}{30}\right)
\end{equation}
This relation is based on the same Salpeter Initial Mass Function as we
have used in deriving our volume-averaged star formation rate.  $C$ is
the concentration factor of neutral hydrogen, $C=\left< \rho^{2}_{\rm
HI}\right> \left< \rho_{\rm HI}\right> ^{-2}$. Simulations suggest
$C\approx 30$ (Gnedin \& Ostriker 1997). Our comoving
star formation rate of $0.005\,h_{70}\,M_{\odot}\,{\rm yr}^{-1}\,{\rm
Mpc}^{-3}$ from the $i'$-drop galaxies we detect is a factor of $>2.5$
lower than the original Madau, Haardt \& Rees (1999) requirement at
$z\approx 5$.  We have updated their equation 27 for the more recent
concordance cosmology estimate of the baryon density of Spergel et al.\
(2003), $\Omega_b=0.0224\,h_{100}^{-2}=0.0457\,h_{70}^{-2}$, and for the
predicted mean redshift of our sample ($z=6.0$):
\begin{equation}
{\dot{\rho}}_{\rm SFR}\approx \frac{0.026\,M_{\odot}\,{\rm yr}^{-1}\,{\rm Mpc}^{-3}}{f_{\rm esc}}\,\left( \frac{1+z}{7}\right) ^{3}\,\left( \frac{\Omega_{b}\,h^2_{70}}{0.0457}\right) ^{2}\,\left( \frac{C}{30}\right)
\end{equation}

The escape fraction of ionizing photons ($f_{\rm esc}$) for
high-redshift galaxies is highly uncertain (e.g., Steidel, Pettini \&
Adelberger 2001), but even if we take $f_{\rm esc}=1$ (no absorption by
H{\scriptsize~I}) this estimate of the star formation density required
is a factor of $\approx 5$ higher than our measured star formation
density of $0.005\,h_{70}\,M_{\odot}\,{\rm yr}^{-1}\,{\rm Mpc}^{-3}$ at
$z\approx 6$ from galaxies in the UDF with
SFRs\,$>1.5\,h_{70}^{-2}\,M_{\odot}\,{\rm yr}^{-1}$.  For faint end
slopes of $\alpha ~-1.8\rightarrow-1.3$ galaxies with $L>0.1\,L^{*}$
account for $32-80$\% of the total luminosity, so would fall short of
the required density of Lyman continuum photons required to reionize the
Universe.  If the faint-end slope is as steep as $\alpha\approx -1.9$
then there would just be enough UV Lyman continuum photons generated in
star forming galaxies at $z\approx 6$ (assuming a Salpeter IMF), but the
requried escape fraction for complete reionization would still have to
be implausibly high ($f_{esc}\approx 1$, whereas all high-$z$
measurements to date indicate $f_{esc}\ll 0.5$: Fern\'{a}nadez-Soto, Lanzetta
\& Chen 2003; Steidel, Adelberger \& Pettini 2001).

AGN are also under-abundant at these epochs (e.g., Dijstra, Haiman \&
Loeb 2004). If star forming galaxies at redshifts close to $z=6$ were
responsible for the bulk of reionization, then a very different initial
mass function would be required, or the calculations of the clumping
factor of neutral gas would have to be significantly over-estimated.
Alternatively another low-luminosity population (e.g., forming globular
clusters; Ricotti 2002) could be invoked to provide some of the
shortfall in ionizing photons.  It is also plausible that the bulk of
reionization occured at redshifts well beyond $z=6$: the WMAP
polarization data indicate  $z_{reion}>10$ (Kogut et al.\ 2003), and it
is possible that the Gunn-Peterson troughs seen in $z> 6.2$ QSOs (Becker
et al.\ 2001; Fan et al.\ 2002) mark the very last period of a neutral
IGM.

\section{Conclusions}
\label{sec:concs}

We summarize our main conclusions as follows:

\begin{enumerate}

\item{We present an $i'$-dropout catalogue of $z\approx 6$ star
forming galaxy candidates in the Ultra Deep Field (UDF) to a
limiting flux ($8\,\sigma$) of $z'_{AB}<28.5$. This represents a
substantial advance over the depths achieved in the GOODS catalogues
and enables us, for the first time, to address questions
concerning the contribution of the faint end of the luminosity
function.}

\item{We detect 54 resolved sources with $(i'-z')_{AB}>1.3$ in the
deepest 11\,arcmin$^{2}$ portion of the UDF and consider this to be an
upper limit to the abundance of star forming galaxies at $z\approx 6$.}


\item{Using simulations based on lower redshift data, we deduce that,
regardless of contamination by foreground interlopers, the abundance of
$i'$-dropouts detected is significantly less than predicted on the basis
of no evolution in the comoving star formation rate from $z=3$ to $z=6$
(integrating to the same luminosity level). The UDF data supports our
previous suggestions that the star formation rate at $z\approx 6$ was
about $\times 6$ {\em less} than at $z\approx 3$ (Stanway, Bunker \&
McMahon 2003).}

\item{The inferred comoving star formation rate of
$0.005\,h_{70}\,M_{\odot}\,{\rm yr}^{-1}\,{\rm Mpc}^{-3}$ from
$L>0.1\,L^{*}_{UV}$ galaxies at $z\approx 6$ may poses a significant
challenge for models which require that luminous star forming galaxies
in the redshift range 6$<z<$10 are responsible for reionizing the
Universe.}

\item{The contamination of our $i'$-drop sample of
candidate $z\approx 6$ galaxies by cool Galactic stars
appears to be minimal at $z'_{AB}>26$, possibly because we
are seeing beyond the Galactic disk at the faint magnitudes probed
by the UDF.}

\end{enumerate}

\subsection*{Note Added in Proof}
A recent preprint by Yan \& Windhorst (astro-ph/0407493)
  independently repeats our selection of candidate $z\approx 6$
  galaxies in the Hubble Ultra Deep Field  with $(i'-z')_{AB}> 1.3$
(astro-ph/0403223 and this paper).
The Yan \& Windhorst catalogue also pushes to fainter magnitudes
than our $z'_{AB}<28.5$ limit, where the completeness corrections become
significant. This subsequent independent analysis recovers almost all of
our original $i'$-band drop-out galaxies, and the catalogues agree at the
98\% level (one discrepant object out of 50). In astro-ph/0407562
(Bunker \& Stanway 2004) we present a matched catalogue of
these $i$-band dropouts in the Hubble Ultra Deep Field.

\subsection*{Acknowledgements}

We thank Steve Beckwith and colleagues at the Space Telescope Science
Institute for making the UDF data available as a public database, on
schedule and in a manner suitable for immediate analysis. ERS
acknowledges a Particle Physics and Astronomy Research Council (PPARC)
studentship supporting this study. Based on observations made with the
NASA/ESA Hubble Space Telescope, obtained from the Data Archive at the
Space Telescope Science Institute, which is operated by the Association
of Universities for Research in Astronomy, Inc., under NASA contract NAS
5-26555. These observations are associated with program \#9978. We
thanks the anonymous referee for some helpful suggestions.

\bsp

\end{document}